\documentclass[11pt,twocolumn]{article}
\usepackage[a4paper, margin=0.75in]{geometry}
\usepackage{amsmath, amssymb, amsthm}
\usepackage{graphicx}

\usepackage{dcolumn}
\usepackage{bm}
\usepackage[utf8]{inputenc}
\usepackage[T1]{fontenc}
\usepackage{mathptmx}
\usepackage{etoolbox}
\usepackage {cite}

\title{Engineering biphoton spectral wavefunction in a silicon micro-ring resonator with split resonances}
\author{Liao Ye,$^{1}$ Haoran Ma,$^{1}$ Xiaoqing Guo,$^{1}$ Fanjie Ruan,$^{1}$ 
	\\Yuehai Wang,$^{1}$ and Jianyi Yang$^{1,*}$ 
	\\Institute of Microelectronic and Nanoelectronics, 
	\\College of Information Science and Electronics Engineering, 
	\\Zhejiang University, Hangzhou 310027, China
	\\12031018@zju.edu.cn}
\date{\today}

\begin{document}
\begin{sloppypar}
	\maketitle
	
	\begin{abstract}
		Frequency-time is a degree of freedom suitable for photonic high-dimensional entanglement, with advantages such as compatibility with single-mode devices and insensitivity to dispersion. The engineering control of the frequency-time amplitude of a photon's electric field has been demonstrated on platforms with second-order optical nonlinearity. For integrated photonic platforms with only third-order optical nonlinearity, the engineered generation of the state remains unexplored. Here, we demonstrate a cavity-enhanced photon-pair source on the silicon-on-insulator (SOI) platform that can generate both separable states and controllable entangled states in the frequency domain without post-manipulation. By choosing different resonance combinations and employing on-chip optical field differentiation, we achieve independent control over two functions that affect the joint spectral intensity (JSI) of the state. A semi-analytical model is derived to simulate the biphoton spectral wavefunction in the presence of resonance splitting and pump differentiation, and its parameters can be fully determined through fitting-based parameter extraction from the resonator's measured linear response. The measured spectral purity for the separable state is $95.5\pm 1.2\%$, while the measured JSIs for the entangled states show two- or four-peaked functions in two-dimensional frequency space. The experiments and simulations demonstrate the capacity to manipulate the frequency-domain wavefunction in a silicon-based device, which is promising for applications like quantum information processing using pulsed temporal-mode encoding or long-distance quantum key distribution.
	\end{abstract}
	
	\section{Introduction}
	Recent advances in quantum photonic systems using photonic integrated circuits (PICs) have demonstrated excellent stability, scalability, and programmability\cite{bao2023very,zheng2023multichip,chi2022programmable}, enabling the efficient generation and manipulation of multiphoton\cite{reimer2016generation,llewellyn2020chip} or high-dimensional\cite{chi2022programmable,lu2020three,zheng2023multichip,imany201850,clementi2023programmable,borghi2023reconfigurable} quantum states on a single chip. High-dimensional entangled states, in particular, have received much attention due to their versatile applications and significant advantages. They enhance quantum computation efficiency\cite{chi2022programmable,huang2024demonstration} and exhibit noise-resistant and dense-encoding capabilities in quantum communications\cite{cerf2002security,barreiro2008beating,kumar2014controlling}. In recent years, photonic entangled states based on different degrees of freedom (d.o.f.s), such as path\cite{chi2022programmable,lu2020three,zheng2023multichip}, frequency-time\cite{xiong2015compact,imany201850,clementi2023programmable,borghi2023reconfigurable}, and transverse spatial modes\cite{feng2022transverse}, have been proposed and experimentally demonstrated on PICs.
	
	Frequency-time, as a potential on-chip d.o.f., enables the encoding of quantum information in high-dimensional Hilbert spaces within specific spatial structures, such as single-mode photonic integrated devices\cite{Ansari:18} and single-mode optical fibers. Many recent on-chip frequency-time schemes concentrate on creating and manipulating frequency-bin\cite{imany201850,clementi2023programmable,borghi2023reconfigurable} or time-bin\cite{xiong2015compact} states. However, due to the discrete nature of bins, these states have a finite number of encodable dimensions within specific frequency or time ranges. Another potential method is to encode quantum information in the complex frequency-time amplitude of a single photon's electric field\cite{brecht2015photon}, which theoretically offers arbitrary encodable dimensions within a specific frequency range\cite{Ansari:18} and could reach the continuous variable (CV) quantum information regime at high dimensionality\cite{eckstein2011highly,brecht2013characterizing,boucher2015toolbox,francesconi2020engineering}. The controlled generation of these states has been extensively demonstrated using parametric down-conversion (PDC) on various second-order optical nonlinear platforms\cite{eckstein2011highly,ansari2018tomography,ansari2018heralded,francesconi2020engineering,jin2021quantum,jin2022two}. To our knowledge, no experiment has yet demonstrated the controlled generation of such frequency-time states on third-order optical nonlinear platforms, such as silicon-on-insulator (SOI). Adapting this scheme to the SOI platform offers significant advantages, including the use of well-developed photonic arbitrary waveform generating\cite{liao2015arbitrary,liao2016photonic} or signal processing\cite{liu2016fully} devices to engineer the spectral or temporal amplitude and phase of pump pulses. This allows for direct shaping of biphoton joint spectral amplitude (JSA) at the generation stage, avoiding post-manipulation that reduces the source's brightness \cite{ansari2018heralded,francesconi2020engineering} or the use of 4-f pulse shapers\cite{ansari2018tomography,ansari2018heralded,francesconi2020engineering,danailov1989time}, which weakens integrability. 
	
	In single-mode silicon waveguides, the interacting optical fields in spontaneous four-wave mixing (SFWM) have wavelengths that differ by only a few to several tens of nanometers\cite{silverstone2014chip,Burridge:23}, and their velocities satisfy the symmetric group velocity matching (sGVM) condition\cite{Ansari:18}. This results in broad phase-matching bands near the pump wavelength\cite{paesani2020near}. However, the controlled generation of biphoton entangled or separable states needs a joint spectral distribution in narrow frequency ranges for both signal and idler photons, as demonstrated in the schemes based on the type-\uppercase\expandafter{\romannumeral2} PDC process\cite{eckstein2011highly,ansari2018tomography,ansari2018heralded,francesconi2020engineering}. Thus, it is necessary to introduce additional devices, such as micro-ring resonators (MRRs), to implement direct filtering to constrain broadband phase-matching at the generation stage, as well as to achieve high source brightness due to their field enhancement. MRRs have preliminarily been used as heralded single-photon sources with high spectral purity\cite{liu2020high,paesani2020near,Burridge:23}. In addition, while resonance splitting is employed to regulate the field enhancement function of MRRs, its impact on the spectral wavefunction of photon pairs has only been theoretically investigated\cite{mccutcheon2021backscattering} and not experimentally validated.
	
	In this work, we demonstrate a cavity-enhanced photon-pair source on the SOI platform that exhibits flexibility in engineering biphoton spectral wavefunctions without post-manipulation. The device is based on an all-pass MRR with split resonances due to backscattering. The experiment begins with the generation of a nearly separable biphoton state in the frequency domain, followed by the sequential adjustment of the predefined two-dimensional signal-idler (TDSI) and antidiagonal pump (ADP) functions to generate two types of biphoton entangled states, respectively. The TDSI function is controlled by adjusting the pump's central wavelength to choose different resonance combinations, while the ADP function is controlled by on-chip differentiation of the pump field (via another MRR-based optical temporal differentiator). Additionally, the biphoton spectral wavefunction for SFWM in the presence of intra-cavity backward propagating modes, as well as analytical expressions for the TDSI and ADP functions, are derived using temporal coupled-mode theory (TCMT). The parameters characterizing the forward and backward propagating modes for specific resonances are obtained from the measured spectra using the fitting method, ensuring the accuracy and reliability of the simulations. The experiments and simulations demonstrate a general method for frequency state preparation on integrated photonic platforms with only third-order optical nonlinearity.

	\section{Theoretical Framework}
	\subsection{Adjustment of Biphoton Spectral Wavefunction with Split Resonances}
	Using a standard perturbative approach and neglecting multipair generation, the quantum state generated by the SFWM process in a waveguide with third-order optical nonlinearity can be expressed as\cite{garay2007photon}
	\begin{equation}
	\label{Eq:1}
	\left| \psi  \right\rangle ={{\left| 0 \right\rangle }_{s}}{{\left| 0 \right\rangle }_{i}}+\beta \iint{d{{\omega }_{s}}d{{\omega }_{i}}F({{\omega }_{s}},{{\omega }_{i}})}\hat{a}_{s}^{\dagger }\left( {{\omega }_{s}} \right)\hat{a}_{i}^{\dagger }\left( {{\omega }_{i}} \right){{\left| 0 \right\rangle }_{s}}{{\left| 0 \right\rangle }_{i}},
	\end{equation}
	where $\beta$ represents waveguide length, the material's third-order nonlinear coefficient, and the pump's amplitude. ${\omega}_{s}$ and ${\omega}_{i}$ are the angular frequencies of signal and idler photons, respectively. The JSA function $F\left({{\omega }_{s}},{{\omega }_{i}}\right)$ describes the spectral entanglement characteristics of the signal-idler pairs. For a MRR-based photon source driven by a degenerate pulsed pump with a central frequency ${\omega}_{p}$, $F\left({{\omega }_{s}},{{\omega }_{i}}\right)$ is given by the following equation, considering the energy conservation constraint\cite{vernon2017truly,christensen2018engineering}:
	\begin{eqnarray}
	\label{Eq:2}
	F({{\omega }_{s}},{{\omega }_{i}})=\int{d{{\omega }_{p}}{{\alpha }_{p}}\left( {{\omega }_{p}} \right)}{{l}_{p}}\left( {{\omega }_{p}} \right){{\alpha }_{p}}\left( {{\omega }_{s}}+{{\omega }_{i}}-{{\omega }_{p}} \right)\nonumber\\
	\times{{l}_{p}}\left( {{\omega }_{s}}+{{\omega }_{i}}-{{\omega }_{p}} \right){{\phi}_{PM}}\left({{\omega }_{p}},{{\omega }_{s}},{{\omega }_{i}}\right){{l}_{s}}\left( {{\omega }_{s}} \right){{l}_{i}}\left( {{\omega }_{i}} \right).
	\end{eqnarray}
	Here, ${{\alpha }_{p}}\left( {{\omega}} \right)$ is the spectral envelope of the degenerate pump, and ${{l}_{x}}\left( {\omega } \right)$ is the ratio of intra-cavity fields to input fields (where $x\text{ = p,s,i}$), also known as field enhancement. The phase-matching function ${{\phi}_{PM}}\left({{\omega }_{p}},{{\omega }_{s}},{{\omega }_{i}}\right)$ is determined by the group velocities of the interacting fields under the linear approximation (see Appendix A). For resonances without splitting, ${{l}_{x}}\left( {\omega } \right)$ can be expressed as ${{l}_{x}}\left( \omega  \right)={{\kappa }_{x}}/\left[ i\left( \omega -{{\omega }_{x}} \right)+1/{{\tau }_{x}} \right]$\cite{fan2003temporal}, where ${{\kappa }_{x}}$ is the field coupling coefficient, ${{\omega }_{x}}$ is the resonant frequency, and $1/{{\tau }_{x}}$ is the amplitude decay rate of the intra-cavity field. Considering the resonance splitting occurring in MRR with a high quality factor\cite{bogaerts2012silicon,li2016backscattering}, ${{l}_{x}}$ is altered to ${{l}_{xf}}$ and ${{l}_{xb}}$ to distinguish between forward and backward propagating modes\cite{mccutcheon2021backscattering}. The TCMT-derived model (see Appendix B) describes ${{l}_{xf}}$ and ${{l}_{xb}}$ as 
	\begin{equation}
	\label{Eq:3}
	{{l}_{xf}}\left( \omega  \right)={{\kappa }_{x}}\frac{-{\gamma_x}{{\mu }_{x,12}}+\left( \omega -{{\omega }_{x}} \right)-i/{{\tau }_{x}}}{{{\left[ i\left( \omega -{{\omega }_{x}} \right)+1/{{\tau }_{x}} \right]}^{2}}+{{\mu }_{x,12}}{{\mu }_{x,21}}},
	\end{equation}
	\begin{equation}
	\label{Eq:4}
	{{l}_{xb}}\left( \omega  \right)={{\kappa }_{x}}\frac{-{{\mu }_{x,21}}+{\gamma_x}\left( \omega -{{\omega }_{x}} \right)-i{\gamma_x}/{{\tau }_{x}}}{{{\left[ i\left( \omega -{{\omega }_{x}} \right)+1/{{\tau }_{x}} \right]}^{2}}+{{\mu }_{x,12}}{{\mu }_{x,21}}},
	\end{equation}
	where ${{\kappa }_{x}}$ is modified to the forward field-coupling coefficient, and ${\gamma_x}$ is the ratio of the backward and forward field coupling coefficients in the coupler. ${{\mu }_{x,12}}$ and ${{\mu }_{x,21}}$ are mutual power coupling coefficients within the ring waveguide, similar to the parameters in Ref.~\cite{li2016backscattering}. When resonance splitting is considered, the JSA observed from the output port (forward propagating mode of the bus waveguide) is regarded as the weighted coherent sum of the contributions from four intra-cavity propagating mode pairs of pump-signal/idler resonances\cite{mccutcheon2021backscattering}, including forward-forward (f-f), forward-backward (f-b), backward-forward (b-f), and backward-backward (b-b) pairs. We use the uppercase symbols $L$ and $\Gamma$ to represent scalar arrays containing these contributions and their weights, respectively. By substituting Eqs.~(\ref{Eq:3}) and (\ref{Eq:4}) into Eq.~(\ref{Eq:2}) and traversing the forward and backward propagating modes of the pump and signal-idler, we obtain
	\begin{eqnarray}
	\label{Eq:5}
	F({{\omega }_{s}},{{\omega }_{i}})=\sqrt{N}\int{d{{\omega }_{p}}{{\alpha }_{p}}\left( {{\omega }_{p}} \right)}{{\alpha }_{p}}\left( {{\omega }_{s}}+{{\omega }_{i}}-{{\omega }_{p}} \right)\nonumber\\
	\times{{\phi }_{PM}}\left( {{\omega }_{p}},{{\omega }_{s}},{{\omega }_{i}} \right)\Gamma L,	
	\end{eqnarray}
	where $\Gamma =\left( 1,~~{{\gamma }_{s}}{{\gamma }_{i}},~~\gamma _{p}^{2},~~\gamma _{p}^{2}{{\gamma }_{s}}{{\gamma }_{i}} \right)$, $L={{\left({{l}_{pf}}l_{pf}^{ec}{{l}_{sf}}{{l}_{if}},~~{{l}_{pf}}l_{pf}^{ec}{{l}_{sb}}{{l}_{ib}},~~{{l}_{pb}}l_{pb}^{ec}{{l}_{sf}}{{l}_{if}},~~{{l}_{pb}}l_{pb}^{ec}{{l}_{sb}}{{l}_{ib}} \right)}^{T}}$. $\sqrt{N}$ is the normalization prefactor. $l_{pf}^{ec}$ and $l_{pb}^{ec}$ are the field enhancement terms considering the energy conservation constraint.
	
	\begin{figure}
	\includegraphics[width=1\linewidth]{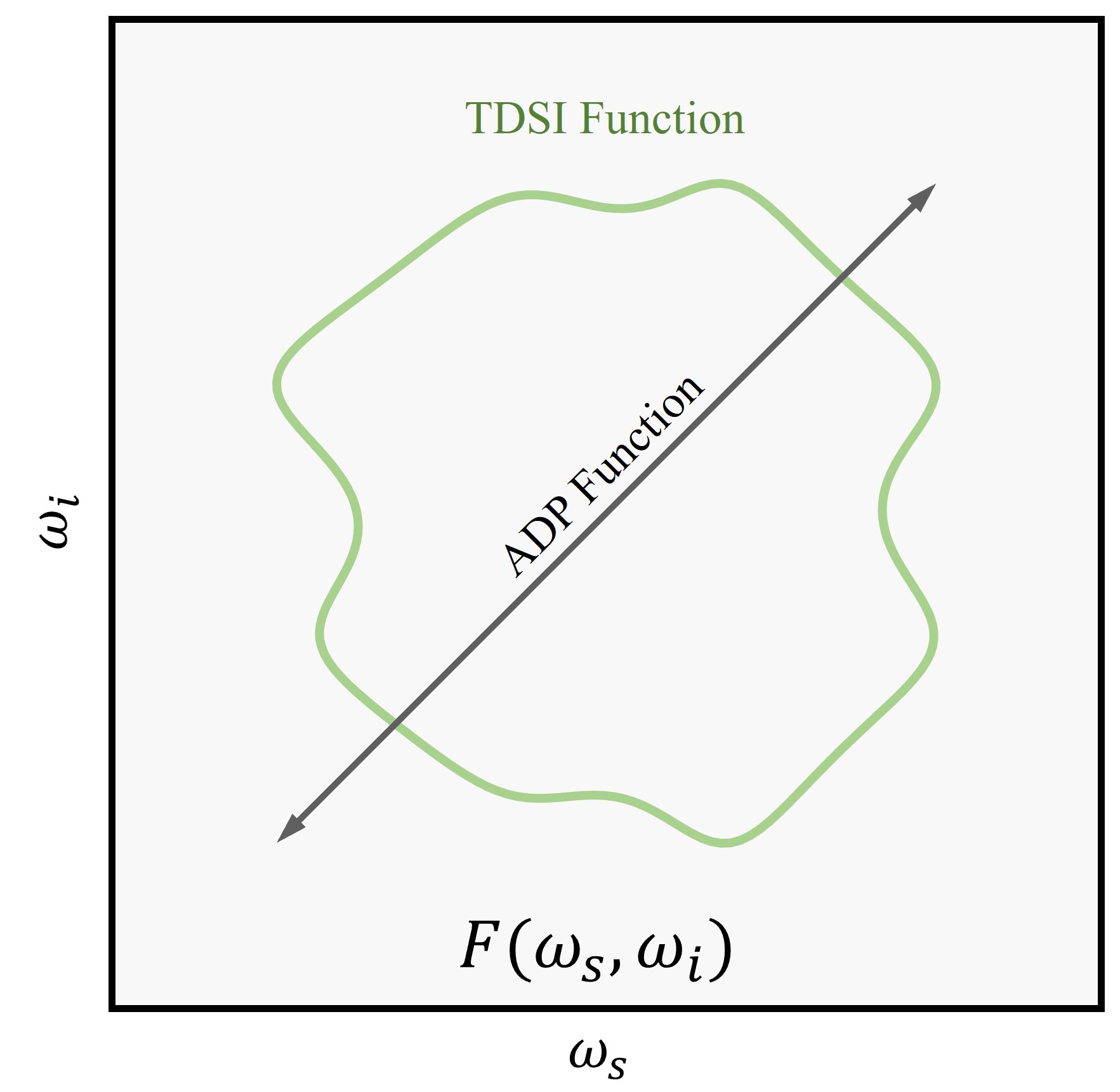}
	\caption{\label{fig:1} Schematic diagram of the biphoton spectral wavefunction engineering principle.}
	\end{figure}
	\subsection{Biphoton Spectral Wavefunction Engineering}
	In order to improve the reliability of simulations, parameters in Eqs.~(\ref{Eq:3}) and (\ref{Eq:4}) for all resonances are determined by fitting the measured transmission spectra (representing the MRR's linear response) to the through-port optical field expression (refer to Appendix B) using a least-squares algorithm. Figure~\ref{fig:1} illustrates the fundamental concept of manipulating the biphoton spectral wavefunction in the SFWM process. From Eq.~(\ref{Eq:5}), we derive two functions that characterize the distribution of $F\left({{\omega }_{s}},{{\omega }_{i}}\right)$ in $\left\{ {{\omega }_{s}},{{\omega }_{i}} \right\}$ space: the ADP function and the TDSI function, which can be expressed as follows:
	\begin{eqnarray}
	\label{Eq:6}
	ADP\left( {{\omega }_{p}} \right)=\left[ {{\alpha }_{p}}\left( {{\omega }_{p}} \right){{l}_{pf}}\left( {{\omega }_{p}} \right) \right]*\left[ {{\alpha }_{p}}\left( {{\omega }_{p}} \right){{l}_{pf}}\left( {{\omega }_{p}} \right) \right]\nonumber\\
	+\gamma _{p}^{2}\left[ {{\alpha }_{p}}\left( {{\omega }_{p}} \right){{l}_{pb}}\left( {{\omega }_{p}} \right) \right]*\left[ {{\alpha }_{p}}\left( {{\omega }_{p}} \right){{l}_{pb}}\left( {{\omega }_{p}} \right) \right],
	\end{eqnarray}
	\begin{equation}
	\label{Eq:7}
	TDSI\left( {{\omega }_{s}},{{\omega }_{i}} \right)={{l}_{sf}}\left( {{\omega }_{s}} \right){{l}_{if}}\left( {{\omega }_{i}} \right)+{{\gamma }_{s}}{{\gamma }_{i}}{{l}_{sb}}\left( {{\omega }_{s}} \right){{l}_{ib}}\left( {{\omega }_{i}} \right).
	\end{equation}
	The ADP function depends on the field enhancement of the pump resonance and the spectrum of the pump pulse, which affects the amplitude and phase spectral distribution of $F\left({{\omega }_{s}},{{\omega }_{i}}\right)$ along the antidiagonal direction. The TDSI function is determined by the field enhancement of the signal-idler resonances and serves as a two-dimensional filtering function in $\left\{ {{\omega }_{s}},{{\omega }_{i}} \right\}$ space. When resonance splitting occurs, both functions can be expressed as weighted sums of contributions from forward and backward propagating modes, as presented in Eqs.~(\ref{Eq:6}) and (\ref{Eq:7}). It is important to emphasize that these two functions can be independently adjusted in experiments. The device is capable of generating nearly separable states when the ADP function is spectrally broadened through pump resonance splitting, as well as controllable entangled states when the TDSI or ADP function is shaped into specific patterns via signal-idler resonance splitting or pump pulse differentiation. The generation of both states and the corresponding engineering methods are demonstrated through simulations and experiments.
	
	\begin{figure}
	\includegraphics[width=1\linewidth]{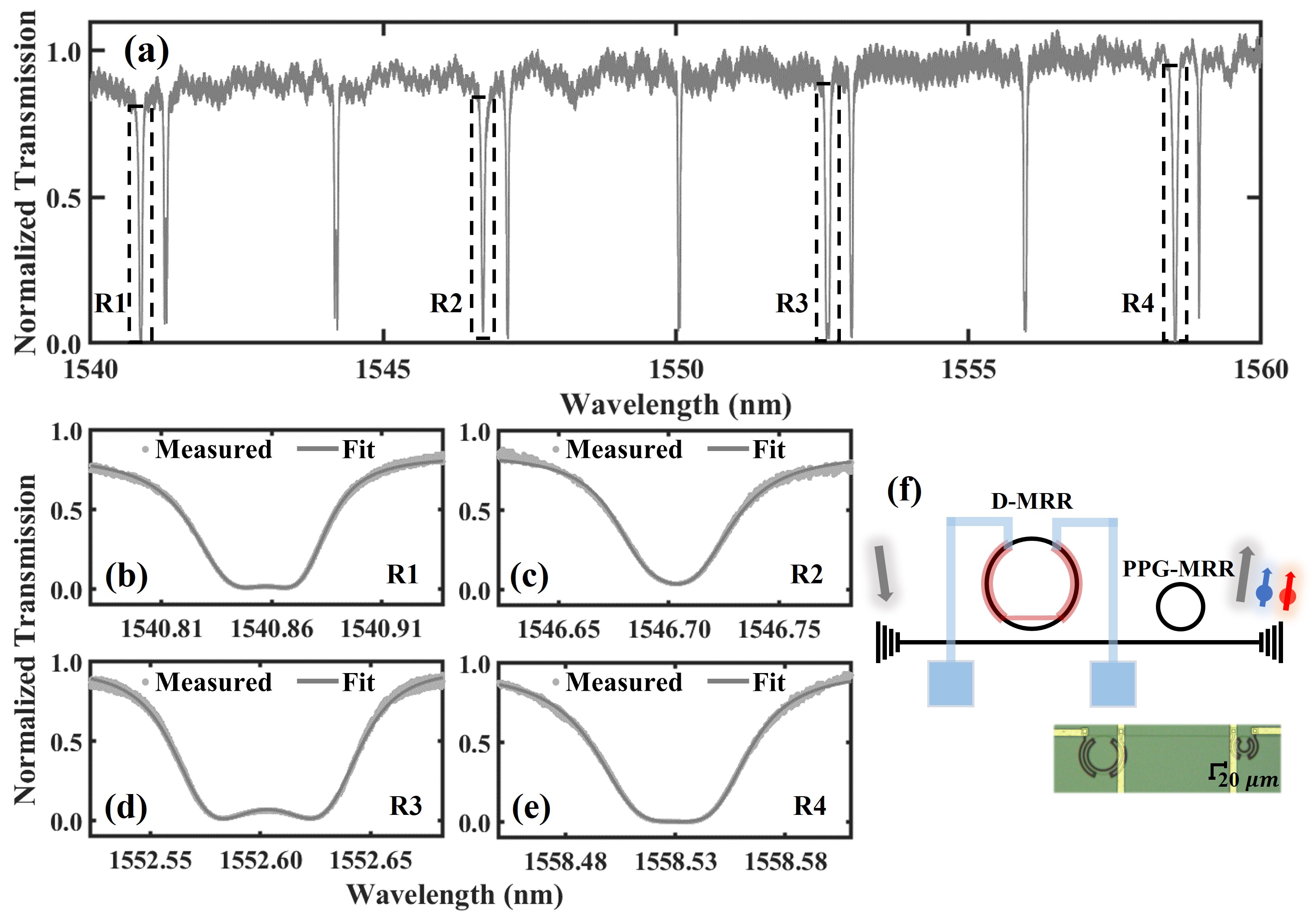}
	\caption{\label{fig:2} Device configurations. (a) Measured spectrum with a 8 volts heater voltage applied to the D-MRR. (b)-(e) Zoomed-in spectra with their fitted curves of R1 (b), R2 (c), R3 (d), and R4 (e). (f) Schematic configuration of the device. Inset: micrograph of the device.}
	\end{figure}
	
	\section{Device configuration}
	A MRR-based photonic temporal differentiator has been demonstrated to perform differentiation operations on optical pulses, producing first- or fractional-order differentiated waveforms\cite{zheng2014fractional}. Figure~\ref{fig:2}(f) shows the schematic configuration of our device, comprising two MRRs indirectly coupled through a single-mode waveguide ($\text{450nm}\times \text{220nm}$ cross-section, TE0 mode): one for optical differentiation (D-MRR) and the other for photon-pair generation (PPG-MRR). Both MRRs operate in the over-coupling regime, with D-MRR and PPG-MRR having radii of 30 $\mu{m}$ and 15 $\mu{m}$, respectively. This results in a free spectral range (FSR) ratio of $1:2$ and a full width at half maximum (FWHM) ratio of approximately $2:5$. A TiN heater is fabricated over D-MRR to tune its resonant wavelengths, enabling the pump's differentiation operation to be switched on or off.
	
	We commence with the regime of generating photons with high spectral purity, which means that the JSA of the generated signal-idler photon pairs is separable. This requires setting the heater voltage to 8 volts, thereby separating the resonances of the two MRRs and deactivating the differentiation operation. Figure~\ref{fig:2}(a) shows the device's spectrum spanning 1540 nm and 1560 nm, with PPG-MRR exhibiting four resonances labeled "R1" through "R4". These are further detailed in Figs.~\ref{fig:2}(b)-\ref{fig:2}(e), illustrating varying degrees of resonance splitting. The free parameters in Eqs.~(\ref{Eq:3}) and (\ref{Eq:4}) for R1-R4 are determined via fitting and listed in Table~\ref{Tb:1} (see Appendix B). To achieve high spectral purity in photon generation, the initial experimental setup selects split R3 as the pump resonance, with relatively unsplit R2 and R4 chosen as the signal and idler resonances, respectively, following the pump-split regime described in Ref.~\cite{mccutcheon2021backscattering}. When referring to the regime of entangled state generation, adjustments in heater voltage and resonance combination are necessary.
	\begin{figure}
	\includegraphics[width=1\linewidth]{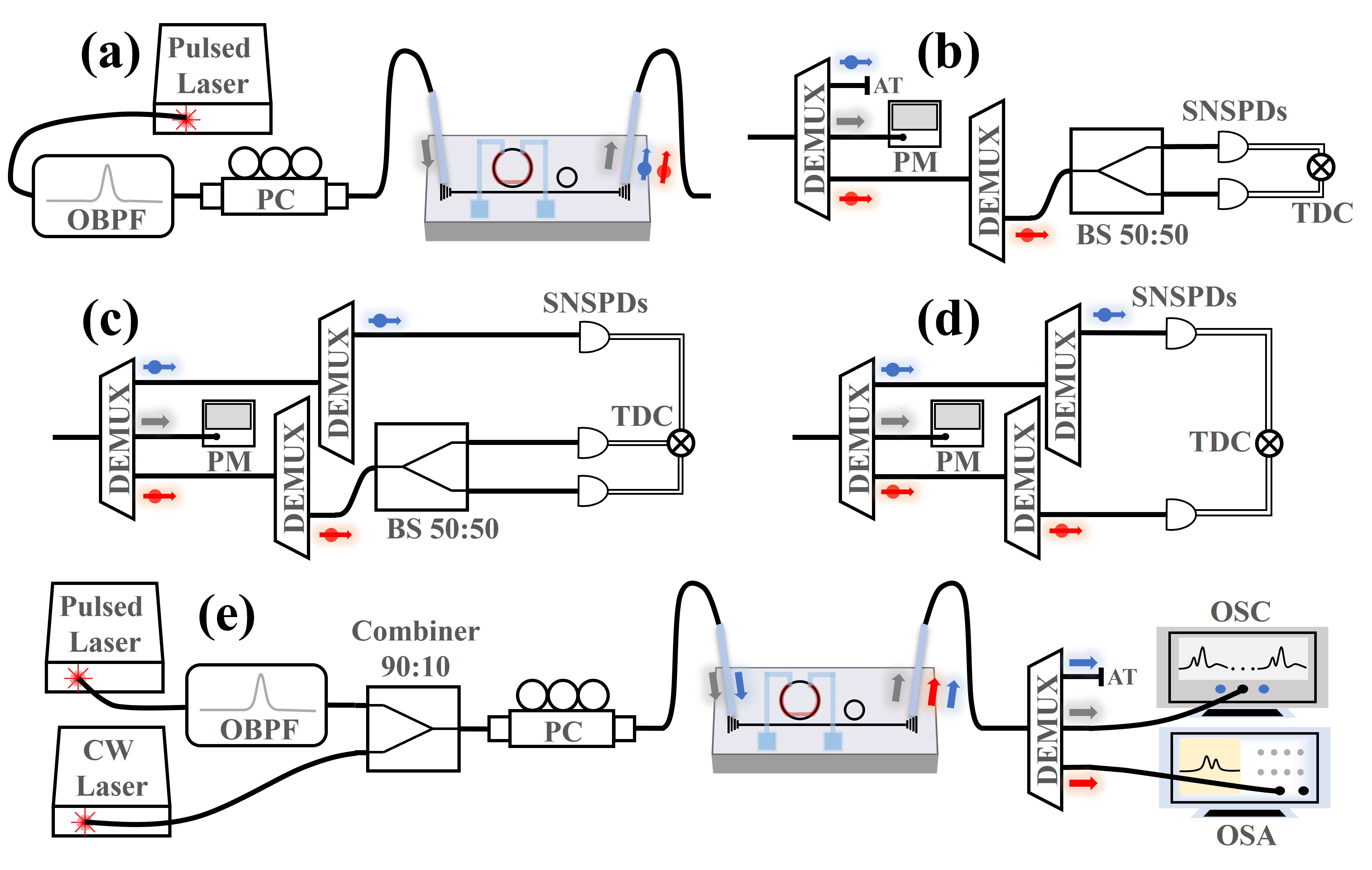}
	\caption{\label{fig:3} Experimental setups. (a) Sketch of the input setup for (b), (c), and (d). OBPF: optical band-pass filter; PC: polarization controller. (b)-(d) Sketches of the output setups for the unheralded second-order correlation (b), conditioned second-order correlation (c), and brightness (d) experiments. DEMUX: demultiplexer; AT: absorbing termination; PM: power meter; BS: beam splitter; SNSPD: superconducting nanowire single-photon detector; TDC: time-to-digital converter. (e) Sketch of the stimulated emission tomography experiment. OSC: oscilloscope; OSA: optical spectrum analyzer.}
	\end{figure}
	
	\section{Experimental setup}
	The experimental setup is shown in Figs.~\ref{fig:3}(a)-\ref{fig:3}(e). Pump pulses are generated by a pulsed erbium-doped fiber laser (Pritel, tunable central wavelength, $\simeq $13.4 ps pulse duration, and 500 MHz repetition rate) and then filtered by a Gaussian-shaped optical band-pass filter (Finisar) with tunable bandwidth, as shown in Fig.~\ref{fig:3}(a). Figure~\ref{fig:3}(b) shows the output setup for unheralded second-order correlation function ($g_{u}^{\left( 2 \right)}\left( \Delta t \right)$) measurement. Two stages of demultiplexers (each with a 200 GHz bandwidth and > 85 dB adjacent channel isolation) are used exclusively for pump rejection, without additional off-chip filters. When the device is used to generate photons with high spectral purity, a significant application is as a heralded single-photon source (SPS)\cite{liu2020high,paesani2020near,Burridge:23}. Therefore, to evaluate its performance, measurements for photon-number purity (determined by the conditioned second-order correlation function $g_{h}^{\left( 2 \right)}\left( \Delta t \right)$) and brightness are added to the output setup, as sketched in Figs.~\ref{fig:3}(c) and~\ref{fig:3}(d), respectively.
	
	Besides, the stimulated emission tomography (SET) technique is used to reconstruct the joint spectral intensity (JSI)\cite{liscidini2013stimulated,eckstein2014high}. Another relatively weak seed beam, generated by a continuous-wave (CW) laser with constant output power, is combined with the pulsed pump beam using a 90:10 beam combiner (where $90\%$ for pump and $10\%$ for seed to mitigate parasitic nonlinear effects), as shown in Fig.~\ref{fig:3}(e). To reconstruct the signal-idler JSI profiles, the idler spectra corresponding to different signal (seed) wavelengths are combined to form a joint spectrum in $\left\{ {{\Delta\lambda }_{s}},{{\Delta\lambda }_{i}} \right\}$ space, where ${\Delta\lambda}_{s}$ and ${\Delta\lambda}_{i}$ represent detunings between the carrier and its neighboring resonant wavelengths.
	\begin{figure*}
	\includegraphics[width=1\linewidth]{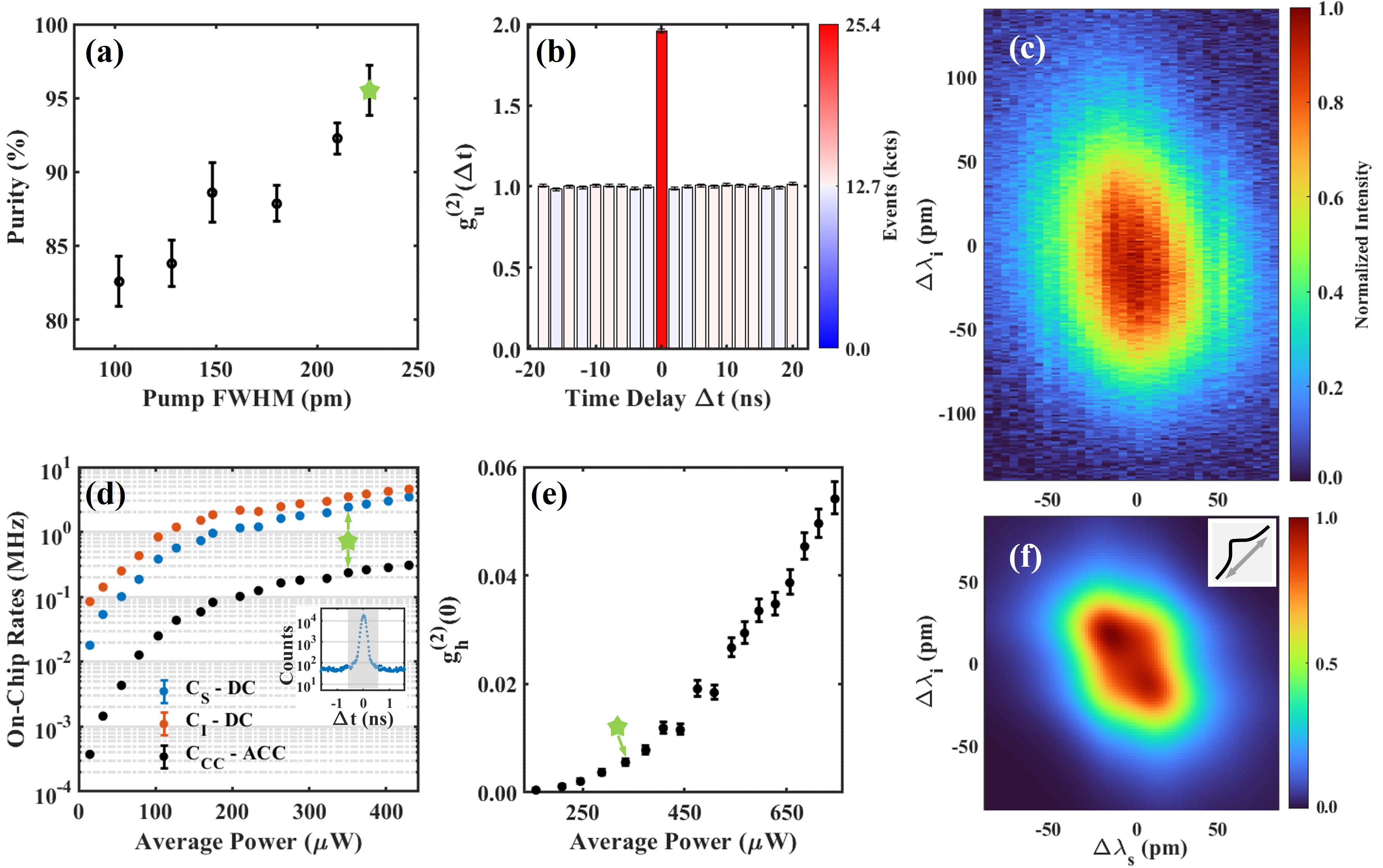}
	\caption{\label{fig:4} (a) Measured purity as a function of the spectral FWHM of the input pump pulses. Each data point corresponds to an unheralded $g_{u}^{(2)}\left( \Delta t \right)$ measurement. (b) Measured unheralded $g_{u}^{(2)}\left( \Delta t \right)$ histograms with delay for the starred data point in (a). Each histogram corresponds to a 660 ps coincidence window (33 bins, each with a 20 ps bin width). (c) Measured joint spectral intensity for the starred data point in (a), consisting of data points in a 36 ($\Delta {{\lambda }_{s}}$-axis) by 282 ($\Delta {{\lambda }_{i}}$-axis) grid. (d) Measured on-chip rates for signal, idler, and signal-idler coincidence as functions of input average power. The dark count rates (DC) and accidental coincidence rates (ACC) have been removed. Inset: raw coincidence counts ${N}_{si}$ as a function of time delay at the average power corresponding to the green star with a 120 s integration time. The shaded area indicates a 940 ps coincidence window for the ${C}_{CC}$ calculation. (e) Measured heralded $g_{h}^{(2)}\left( 0 \right)$ as a function of input average power. Error bars in (a), (b), (d), and (e) represent 1 standard deviation under Poissonian statistics. (f) Modulus squared of the simulated JSA corresponding to (c). Inset: sketch of its ADP function. The same wavelength scales are adopted in (c) and (f).}
	\end{figure*}
	\begin{figure}
	\includegraphics[width=1\linewidth]{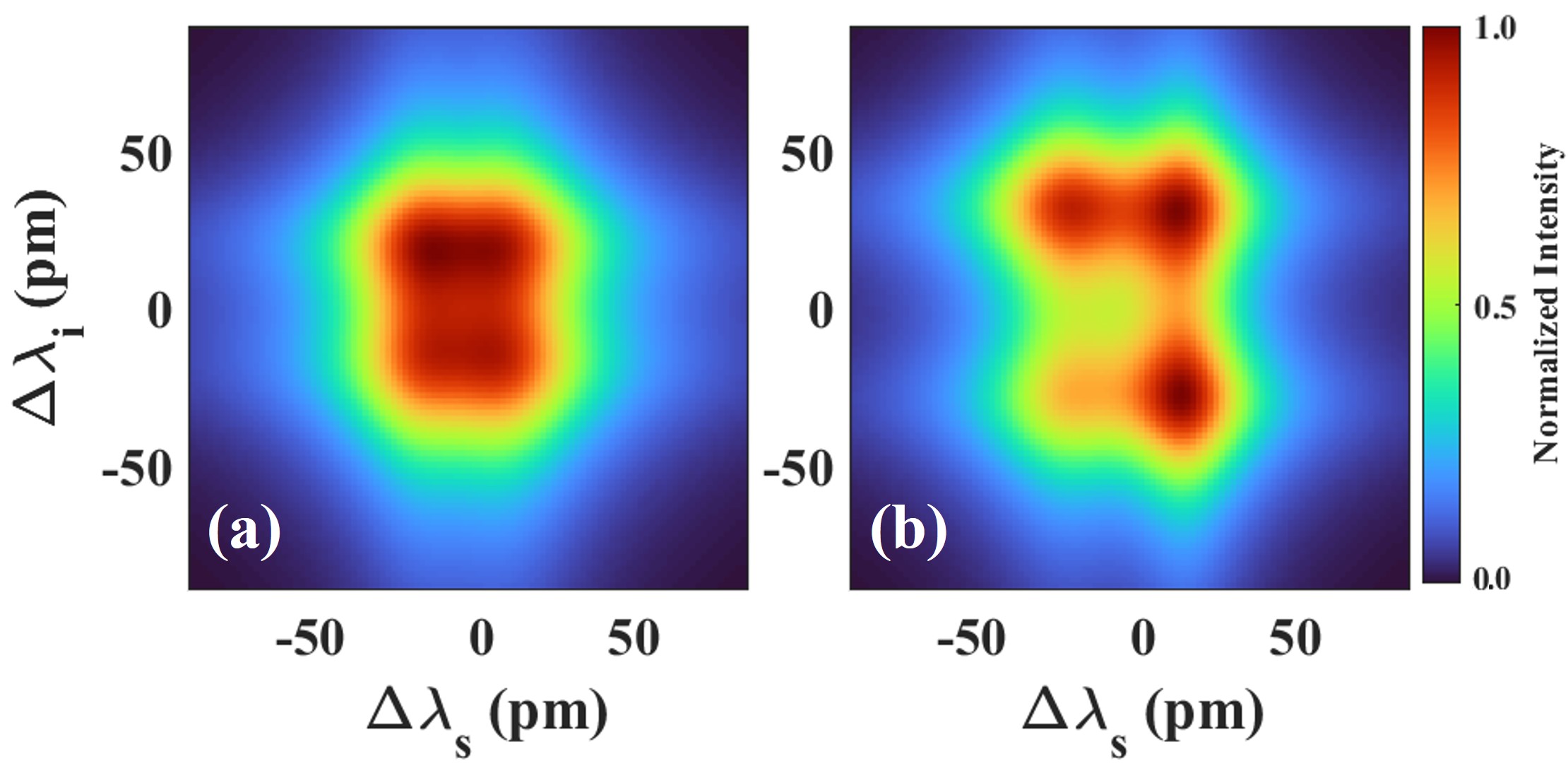}
	\caption{\label{fig:5} Simulated TDSI functions for the separable state generation regime (a) and the entangled state generation regime (b).}
	\end{figure}
	
	\section{Separable state generation}
\begin{figure*}
	\includegraphics[width=1\linewidth]{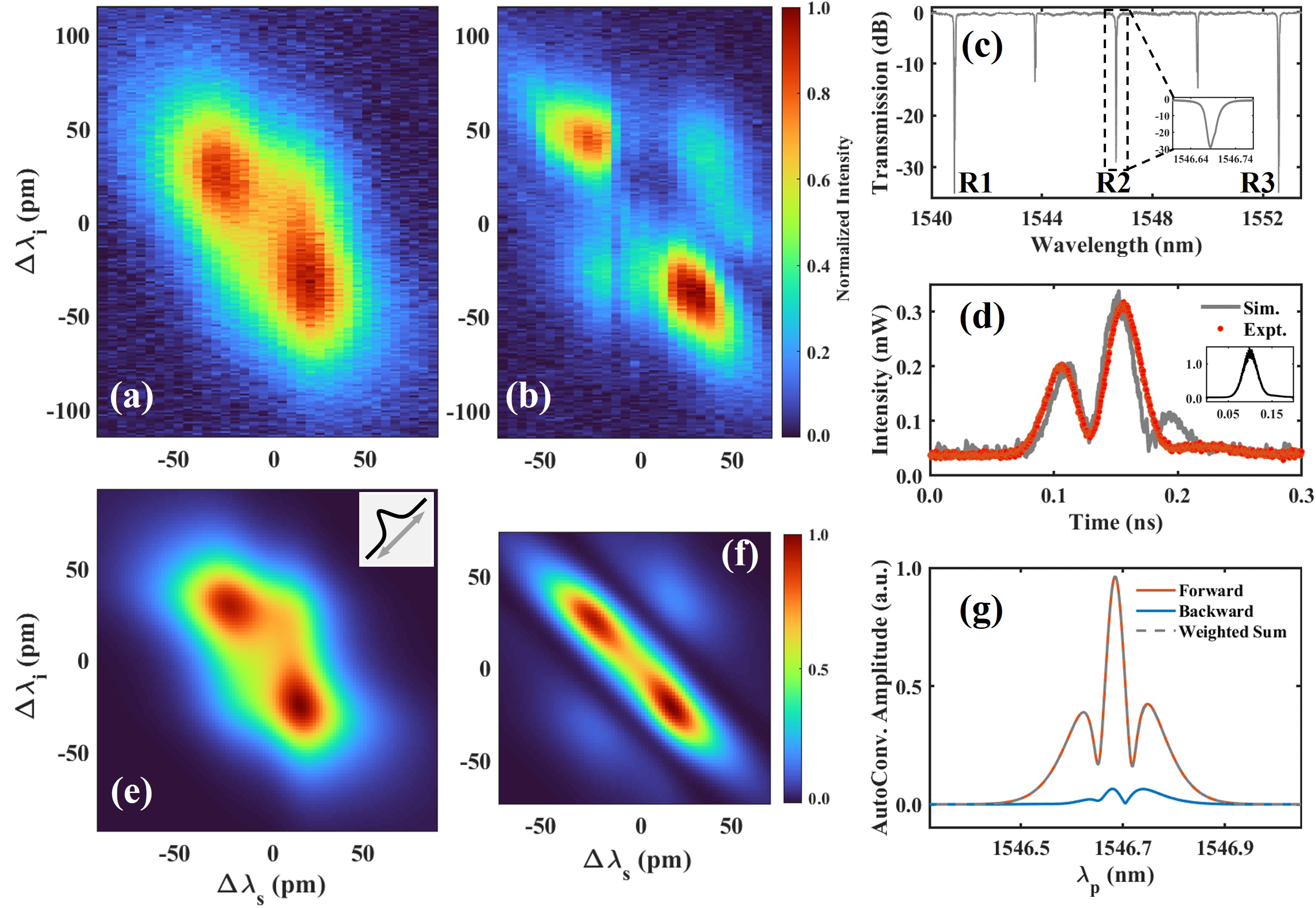}
	\caption{\label{fig:6} (a) Measured JSI with only the TDSI function engineered, consisting of data points in a 37 by 230 grid. (b) Measured JSI with both the TDSI and ADP functions engineered, consisting of data points in a 30 by 230 grid. (c) Measured spectrum with a 5.1-volt heater voltage applied to the D-MRR. (d) Waveforms of the output pump by simulation and experiment. Inset: waveform of the input pump. The oscilloscope's sample frequency is set to 40 GHz, and the averaging time is 64 for the measured waveforms. (e) Modulus squared of the simulated JSA corresponding to (a). Inset: sketch of its ADP function. (f) Modulus squared of the simulated JSA corresponding to (b). The same wavelength scales are adopted in (a), (b), (e), and (f). (g) Simulated auto-convolutions of ${{\alpha }_{p}}\left( {{\omega }_{p}} \right){{l }_{pf}}\left( {{\omega }_{p}} \right)$ (forward) and ${{\alpha }_{p}}\left( {{\omega }_{p}} \right){{l }_{pb}}\left( {{\omega }_{p}} \right)$ (backward), and the ADP function (weighted sum) for the JSI in (f).}
	\end{figure*}

	In the regime of separable state generation, the pump operates at a central wavelength of 1552.61 nm (channel 31 of the C-band ITU grid), while the signal and idler have central wavelengths of 1546.70 nm and 1558.52 nm, respectively. We measure the time-integrated $g_{u}^{\left( 2 \right)}\left( \Delta t \right)$ of idler photons at various pump spectral FWHM. In the low squeezing regime, the spectral purity $P$ correlates with the zero-delay second-order correlation function $g_{u}^{\left( 2 \right)}\left( 0 \right)$, as described by $P={{g}_{u}^{(2)}}(0)-1$\cite{christ2011probing,signorini2020chip}. Figure~\ref{fig:4}(b) shows the measured $g_{u}^{\left( 2 \right)}\left( \Delta t \right)$ with the pump's FWHM set to 226 pm. The measured $g_{u}^{\left( 2 \right)}\left( 0 \right)$ is $1.955\pm 0.012$, corresponding to a spectral purity of $95.5\pm 1.2\%$, which approximates the $97\%$ upper limit predicted for a MRR-based SPS with only pump splitting\cite{mccutcheon2021backscattering}. This purity surpasses the theoretical $\sim93\%$ upper limit of an MRR-based SPS\cite{helt2010spontaneous,liu2020high} without additional purity-enhancement designs. The measured JSI from the SET experiment is shown in Fig.~\ref{fig:4}(c) with the same 226 pm FWHM pump input. In the context of discrete-variable description\cite{brecht2013characterizing}, the entanglement between signal and idler photons is extracted by the Schmidt decomposition of the JSA function. Assuming a flat-phase JSA\cite{burridge2020high,Burridge:23}, the Schmidt decomposition of the square root of the measured JSI yields a spectral purity of $97.2\%$. Although the low spectral resolution (0.02 nm) of the OSA affects the distribution of JSI, causing noticeable broadening along the $\Delta\lambda_{i}$ axis, the purity value remains reasonably consistent with the $g_{u}^{\left( 2 \right)}\left( \Delta t \right)$ measurements. We are more concerned with the reason for the spectral purity exceeding $93\%$ in the presence of pump resonance splitting. Figure~\ref{fig:4}(f) shows the numerical simulation results for the JSI and its ADP function. Pump resonance splitting significantly broadens the ADP function (inset) compared to the case without pump splitting (inset of Fig.~\ref{fig:6}(e)). This broadening of the ADP function leads to a corresponding broadening of the JSI distribution along the antidiagonal direction, thereby improving pectral purity. When the pump FWHM is gradually reduced from 226 pm, the purity measured by the $g_{u}^{\left( 2 \right)}\left( \Delta t \right)$ experiment exhibits an overall decreasing trend, as shown in Fig.~\ref{fig:4}(a). We also measure several corresponding JSIs and compare them to numerical simulations as the FWHM decreases, as shown in Appendix C.
	
	To demonstrate the device's capability as a heralded SPS, we measure both its photon-pair generation rate and photon-number purity. In Fig.~\ref{fig:4}(d), the data points marked by the green star represent the on-chip rates when the average power at the bus's input port is set to the same value as in the previously discussed $g_{u}^{\left( 2 \right)}\left( \Delta t \right)$ experiment. After removing losses from the output grating coupler, two DEMUXs, and SNSPDs (provided in Appendix C), the estimated on-chip photon pair generation rate at this average power is $236.2\pm 0.5$ kHz. The coincidence-to-accidental ratio (CAR) is $62.8\pm 0.2$, as shown in the inset of Fig.~\ref{fig:4}(d). At the same average power, the $g_{h}^{\left( 2 \right)}\left( 0 \right)$ value is $0.0055\pm 0.0006$, as shown in Fig.~\ref{fig:4}(e), demonstrating excellent photon-number purity for our device as a heralded SPS\cite{signorini2020chip}. 
	
	\section{Entangled state generation}
	In the previous regime, the TDSI function exhibits a single-peak characteristic, as shown in Fig.~\ref{fig:5} (a), ensuring state separability. In contrast, to generate biphoton entangled states, the TDSI function needs to be adjusted. We set the pump's central wavelength to 1546.70 nm, which aligns with the central wavelength of the unsplit R2. The relatively split R1 and R3 are chosen as the signal and idler resonances, respectively, resulting in a four-peaked TDSI function, as shown in Fig.~\ref{fig:5} (b). Besides, the pump's FWHM is reduced to 100 pm to enhance the spectral anti-correlation between photon pairs. 
	
	Figure~\ref{fig:6}(a) shows the JSI measured using the SET technique with the Gaussian-shaped pump, which exhibits a spectral distribution with two peaks along the main diagonal direction. The corresponding simulated JSI is shown in Fig.~\ref{fig:6}(e). The other two peaks of the TDSI function along the antidiagonal direction are absent in both the measured and simulated JSI. This absence is attributed to the tight energy conservation constraints along the antidiagonal direction, caused by the unsplit pump resonance and the narrow bandwidth of the input pump pulse. This constraint is consistent with the narrow bandwidth of the simulated ADP function, as shown in the inset of Fig.~\ref{fig:6}(e).
	
	To further engineer the spectral distribution along the antidiagonal direction, the heater voltage is set to 5.1 volts, causing the pump resonance of the PPG-MRR to spectrally overlap with the adjacent resonance of the D-MRR, as shown in Fig.~\ref{fig:6}(c). As a result, the pump is filtered with the D-MRR's transfer function (see Eq.~\ref{Eq:C1} in Appendix C), which approximates an ideal differentiator's transfer function within a narrow frequency range centered on the resonant frequency\cite{zheng2014fractional}. The differentiated pump pulses are then coupled into the PPG-MRR, generating photon pairs via SFWM. Figure~\ref{fig:6}(d) shows the output pulse after passing through two MRRs, monitored by a wide-bandwidth oscilloscope, which approximates 1.7-order mathematical differentiation of the input pulse (inset). Figure~\ref{fig:6}(b) shows the measured JSI for photon pairs generated by the differentiated pump. The discontinuity along the $\Delta\lambda_{s}$ axis is caused by the instantaneous decrease in the thermo-optic redshift of all resonances when the sweeping seed wavelength crosses the resonant wavelength of R1\cite{de2019power}. The simulation results shown in Fig.~\ref{fig:6}(f) exclude the discontinuity caused by the therm-optic dispersion effect from the SET experiment, resulting in a clearer four-peaked JSI function and demonstrating wavefunction engineering along the antidiagonal direction. While maintaining the two-peaked feature along the main diagonal, the pump differentiation alters the ADP function, which is the weighted sum of two auto-convolutions contributed by the forward and backward propagating modes, respectively, as shown in Fig.~\ref{fig:6}(g) and Eq.~(\ref{Eq:6}). The ADP function (dashed-gray curve) exhibits a three-peaked feature in agreement with the simulated JSI's antidiagonal distribution. 
	
	\section{Discussion and Conclusion}
	While the current device has limitations in generating a broad range of biphoton states, there is considerable potential for refinement in its programmability. By leveraging existing engineering methods, coupled resonator systems could be introduced to enable controllable resonance splitting\cite{zhang2019electronically,okawachi2021dynamic}, and arbitrary waveform generators based on finite impulse response filters\cite{liao2015arbitrary} or the Taylor synthesis method\cite{liao2016photonic} could be employed for pump pulse shaping. These enhancements, which are fully compatible with the SOI platform, could substantially improve the source's programmability. Besides, we note that the biphoton states presented in Figs.~\ref{fig:6}(a) and~\ref{fig:6}(e), following Schmidt decomposition, exhibit weights of approximately 0.91 and 0.09 for the first two Schmidt modes, respectively. This demonstrates the device's capability to generate two-dimensional entangled states for pulsed temporal-mode (PTM) encoding\cite{Ansari:18}. We believe that with the introduction of highly programmable photonic devices, this method has the potential to facilitate the controlled on-chip generation of high-dimensional entangled states for PTM encoding.
	
	In conclusion, we have demonstrated a cavity-enhanced photon-pair source on the SOI platform that can generate both separable states with $95.5\pm 1.2\%$ spectral purity and entangled states with two- or four-peaked biphoton spectral wavefunction in $\left\{ {{\Delta\lambda }_{s}},{{\Delta\lambda }_{i}} \right\}$ space. By choosing different resonance combinations and employing on-chip optical field differentiation, these states are engineered directly at the generation stage. A semi-analytical model is derived to simulate the biphoton spectral wavefunction in the presence of resonance splitting and pump differentiation. The model's parameters can be fully determined through measurement and fitting-based parameter extraction of the MRR's linear response. To quantify the wavefunction engineering process, we introduce two functions: TDSI and ADP. Control over these functions is consistent with our two independent adjustment methods, as demonstrated by both experiments and simulations. 
	
	Operating within the frequency-time d.o.f., the device maintains spatial single-mode properties while offering higher integration compared to PDC-based schemes and higher brightness compared to post-manipulation schemes. Through theoretical analysis and experimental validation, we have demonstrated the capability to manipulate the frequency-domain wavefunction of SFWM photon pairs on the SOI platform. The device, along with the state engineering method, holds promise for applications in long-distance quantum key distribution due to the insensitivity to polarization-mode dispersion\cite{zhang2008distribution,nunn2013large,shi2019multichannel}, as well as quantum information processing based on PTM encoding\cite{brecht2015photon,Ansari:18}.

\section*{ACKNOWLEDGMENT}
This work was supported by the National Key Research and Development Program of China (Grants No. 2021YFB2800201), the National Natural Science Foundation of China (Grants No. U22A2082), and the Ningbo Science and Technology Program (Grants No. 2023Z073).

\section*{AUTHOR DECLARATIONS}
\subsection*{Conflict of Interest}
The authors have no conflicts to disclose.

\section*{DATA AVAILABILITY}
The data that support the findings of this study are available from the corresponding authors upon reasonable request.

\setcounter{equation}{0}
\renewcommand{\theequation}{A\arabic{equation}}
\section*{APPENDIX A: LINEAR APPROXIMATION FOR PHASE MATCHING FUNCTION}
The phase matching function is given by\cite{garay2007photon}
\begin{eqnarray}
\label{Eq:A1}
{{\phi }_{PM}}\left( {{\omega }_{p}},{{\omega }_{s}},{{\omega }_{i}} \right)= \text{sinc}\left[ \frac{L}{2}\Delta k\left( {{\omega }_{p}},{{\omega }_{s}},{{\omega }_{i}} \right) \right]\nonumber\\
\times\exp \left[ i\frac{L}{2}\Delta k\left( {{\omega }_{p}},{{\omega }_{s}},{{\omega }_{i}} \right) \right],
\end{eqnarray}
where ${L}$ is the length of the waveguide. The phase mismatch factor $\Delta k\left( {{\omega }_{p}},{{\omega }_{s}},{{\omega }_{i}} \right)$ can be defined as
\begin{eqnarray}
\label{Eq:A2}
\Delta k\left( {{\omega }_{p}},{{\omega }_{s}},{{\omega }_{i}} \right)=k\left( {{\omega }_{p}} \right)+k\left( {{\omega }_{s}}+{{\omega }_{i}}-{{\omega }_{p}} \right)\nonumber\\
-k\left( {{\omega }_{s}} \right)-k\left( {{\omega }_{i}} \right)-\gamma P, 
\end{eqnarray}
where $\gamma$ represents the nonlinear parameter including both self-phase and cross-phase modulation, and $P$ is the peak power of the incident pulsed pump. $k\left( {{\omega }_{p}}\right)$, $k\left( {{\omega }_{s}}+{{\omega }_{i}}-{{\omega }_{p}} \right)$,  $k\left( {{\omega }_{s}}\right)$, and $k\left( {{\omega }_{i}}\right)$ are wavevectors of four interacting fields, which are expanded in first-order Taylor series at perfectly phase-matched frequencies $\omega _{x}^{0}$ (where ${x}=p,s,i$ for the degenerate pump regime). The coefficients for the expansion are expressed as $k_{x}^{\left( n \right)}\left( \omega  \right)={{d}^{n}}{{k}_{x}}/d{{\omega }^{n}}{{|}_{\omega =\omega _{x}^{0}}}$. Therefore, the term $\Delta {{k}^{(0)}}\approx 2{{k}^{\left( 0 \right)}}\left( \omega _{p}^{0} \right)-{{k}^{\left( 0 \right)}}\left( \omega _{s}^{0} \right)-{{k}^{\left( 0 \right)}}\left( \omega _{i}^{0} \right)-\gamma {{P}_{0}}$ vanishes at the given ${{P}_{0}}$, and the linear approximation for Eq.~(\ref{Eq:A2}) is derived as\cite{paesani2020near,garay2007photon}
\begin{equation}
\label{Eq:A3}
\Delta {{k}_{lin}}={{\tau }_{s}}{{\nu }_{s}}+{{\tau }_{i}}{{\nu }_{i}},
\end{equation}
where  ${{\nu }_{s}}={{\omega }_{s}}-\omega _{s}^{0}$ and ${{\nu }_{i}}={{\omega }_{i}}-\omega _{i}^{0}$ are angular-frequency detunings. ${{\tau }_{s}}=k_{p}^{\left( 1 \right)}(\omega _{p}^{0})-k_{s}^{\left( 1 \right)}(\omega _{s}^{0})$ and ${{\tau }_{i}}=k_{p}^{\left( 1 \right)}(\omega _{p}^{0})-k_{i}^{\left( 1 \right)}(\omega _{i}^{0})$ are group-velocity mismatches.
\begin{figure}
\includegraphics[width=1\linewidth]{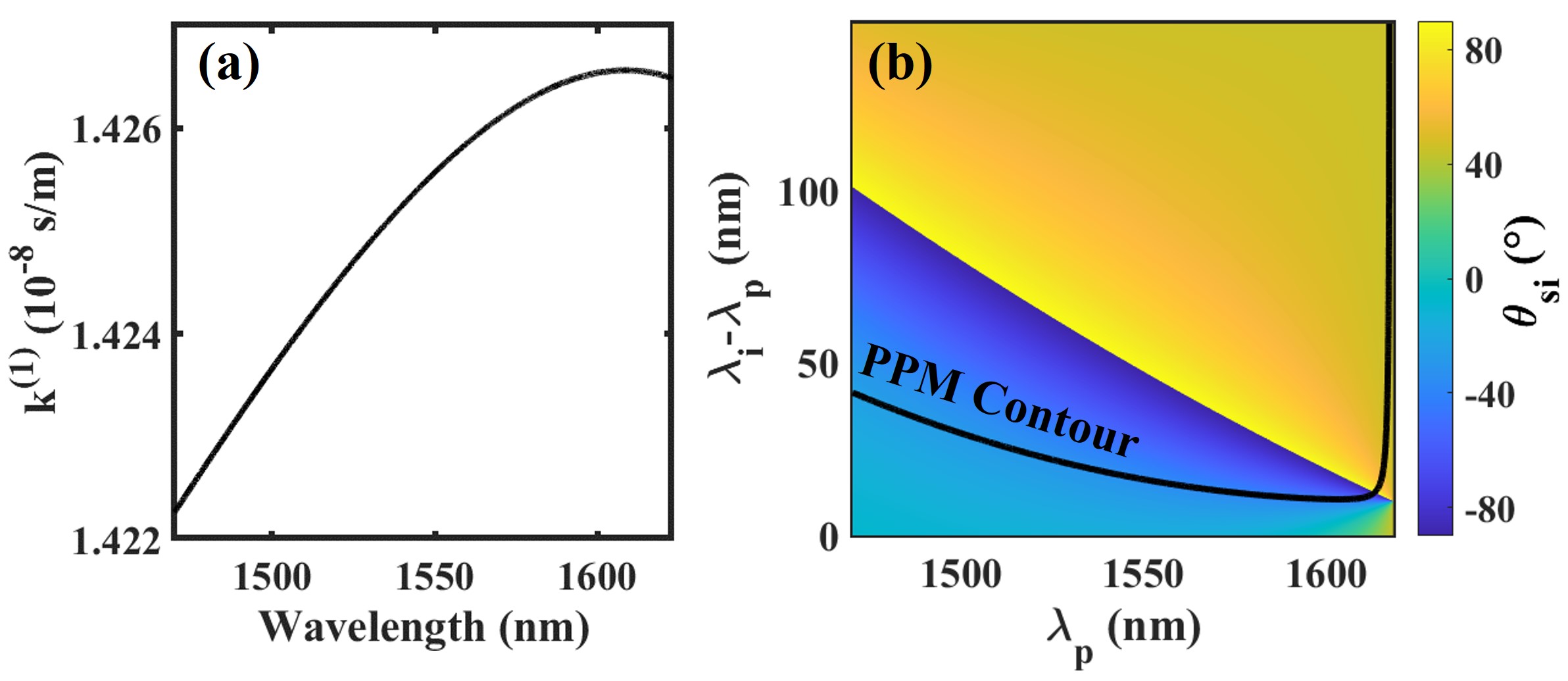}
\caption{\label{fig:7} (a) Simulated first-order Taylor expansion coefficient in wavelength. (b) Simulated phase-matching orientation angle in $\left\{ {{\lambda }_{p}},{{\lambda }_{i}}-{{\lambda }_{p}} \right\}$ space. The black curve indicates the perfect phase-matching contour.}
\end{figure}
\begin{table*}
\centering
\caption{\label{Tb:1}The parameters of the four resonances located within the wavelength range of 1540 nm to 1560 nm of the PPG-MRR.}
\begin{tabular}{ 
c c c c c c c c c c
}
\hline\hline\\
Resonance & C & $1/\tau$ (THz) & $\gamma$ & $\mu_0$ (THz) & $\kappa$ ($\sqrt{\text{THz}}$) & $\phi_1$ & $\phi_2$ & $\omega_0$ (THz) & R-squared \\[0.2cm] \hline\\
R1 & 0.920 & 0.0167 & 0.389 & 0.0152 & 0.145 & 1.990 & 4.614 & 1222.47 & 0.999 \\
[0.2cm] 
R2 & 0.930 & 0.0183 & 0.291 & 0.0114 & 0.159 & 0.793 & 5.262 & 1217.85 & 0.998 \\
[0.2cm]
R3 & 0.976 & 0.0190 & 0.572 & 0.0245 & 0.167 & 1.461 & 4.842 & 1213.23 & 0.999 \\
[0.2cm]
R4 & 0.982 & 0.0200 & 0.382 & 0.0161 & 0.164 & 1.740 & 4.688 & 1208.61 & 0.999 \\
[0.2cm]
\hline\hline
\end{tabular}
\end{table*}

Figure~\ref{fig:7}(a) shows the simulated $k^{\left( 1 \right)}$ for the fundamental TE mode of a Si waveguide with a $\text{450nm}\times \text{220nm}$ cross-section, obtained from Ansys Lumerical MODE's dispersion calculations. The phase-matching orientation angle ${{\theta }_{si}}=-\arctan ({{\tau }_{s}}/{{\tau }_{i}})$ is defined as the angle between the brightest strip and the $\lambda_{s}$-axis for ${{\phi }_{PM}}$ in $\left\{ {{\lambda }_{s}},{{\lambda }_{i}} \right\}$ space, illustrating the spectral correlation of ${{\phi }_{PM}}$\cite{garay2007photon,sharma2022spectrally}. Figure~\ref{fig:7}(b) shows ${{\theta }_{si}}$ for $\lambda_{p}$ covering S-, C-, and L-band in $\left\{ {{\lambda }_{p}},{{\lambda }_{i}}-{{\lambda }_{p}} \right\}$ space. The $\theta_{si}^{0}$ at the perfect phase-matching (PPM) contour ranges from ${-33.5}^{\circ}$ to ${-36.3}^{\circ}$ with $\lambda_{p}$ from 1540 nm to 1560 nm, covering the wavelength range of the experiments. These $\theta_{si}^{0}$ values approximate the $-45^{\circ}$ pump envelope angle under the energy conservation constraint. This demonstrates the inevitable spectral anti-correlation of biphoton states generated in such a waveguide without MRR-enhancement or post-manipulation. 

\setcounter{equation}{0}
\renewcommand{\theequation}{B\arabic{equation}}
\section*{APPENDIX B: TEMPORAL COUPLED-MODE THEORY FOR MICRO-RING RESONATORS WITH RESONANCE SPLITTING}
The backscattering of a MRR is caused by both the ring waveguide's sidewall roughness and the coupler's abrupt effective index transition\cite{li2016backscattering}. Given these backscatterings, the optical field ${a}$, which is normalized to the optical energy stored in an all-pass MRR based on TCMT, splits into two modes: forward propagating mode ${a}_{f}$ and backward propagating mode ${a}_{b}$. These two split modes of resonance with a central frequency of ${\omega}_{0}$ satisfy
\begin{subequations}
\label{Eq:B1}
\begin{align}
\frac{d{{a}_{f}}}{dt}=i{{\omega }_{0}}{{a}_{f}}-\frac{1}{\tau }{{a}_{f}}-i{{\mu }_{12}}{{a}_{b}}-i{\kappa }{{S}_{i}}, \label{B1a}\\
\frac{d{{a}_{b}}}{dt}=i{{\omega }_{0}}{{a}_{b}}-\frac{1}{\tau }{{a}_{b}}-i{{\mu }_{21}}{{a}_{f}}-i{\kappa }'{{S}_{i}}, \label{B1b}
\end{align}
\end{subequations}
where ${{S}_{i}}$ is the amplitude of the input field. ${\kappa}$ and ${\kappa }'$ are the forward and backward field coupling coefficients of the coupler, respectively. ${{\mu }_{12}}$ and ${{\mu }_{21}}$ are the complex power coupling coefficients between forward and backward propagating modes within the ring waveguide. They have the same amplitude but different phases. By defining ${{l}_{j}}\left( \omega  \right)={{a}_{j}}\left( \omega  \right)/{{S}_{i}}\left( \omega  \right)$ (where ${j}={f,b}$) and solving for them in Eqs.~(\ref{B1a}) and (\ref{B1b}), Eqs.~(\ref{Eq:3}) and (\ref{Eq:4}) can be derived. Besides, the output field ${{S}_{o}}$ at the through port is expressed as
\begin{equation}
\label{Eq:B2}
{{S}_{o}}={{S}_{i}}-i\kappa {{a}_{f}}-i\kappa '{{a}_{b}}.
\end{equation}
To better fit the measured transmission spectra at the through port, a dimensionless factor ${\gamma}={\kappa}'/{\kappa}$ is introduced. The complex ${{\mu }_{12}}$ and ${{\mu }_{21}}$ are replaced by real ${{\mu }_{0}}$, ${\phi }_{1}$, and ${\phi }_{2}$, with ${{\mu }_{12}}={{\mu }_{0}}\exp \left( i{{\phi }_{1}} \right)$ and ${{\mu }_{21}}={{\mu }_{0}}\exp \left( i{{\phi }_{2}} \right)$. The ratio of output to input fields is expressed as
\begin{footnotesize}
\begin{equation}
\label{Eq:B3}
\frac{{{S}_{o}}}{{{S}_{i}}}=C+{{\kappa }^{2}}\frac{i\gamma{{\mu }_{0}}\left( {{e}^{i\phi_1}}+{{e}^{i\phi_2}} \right)-\left( 1+{{\gamma}^{2}} \right)\left[ i\left( \omega -{{\omega }_{0}} \right)+1/\tau  \right]}{{{\left[ i\left( \omega -{{\omega }_{0}} \right)+1/\tau  \right]}^{2}}+\mu _{0}^{2}{\exp \left[ i\left( {\phi_1}+{\phi_2} \right) \right] }},
\end{equation}
\end{footnotesize}
where ${C}$ is a normalized amplitude factor for compensating energy losses that may occur while measuring the spectra. Table~\ref{Tb:1} shows the free parameters in Eq.~(\ref{Eq:B3}) obtained by least-squares fit for the four resonances mentioned in the text, as well as R-squared values, supported by MATLAB's Curve Fitting Toolbox. The high R-squared values ($\ge$ 0.998) demonstrate excellent agreement between the fitted curve using Eq.~(\ref{Eq:B3}) and the square root of the measured spectra. The fitting method always yields both forward and backward propagating modes for resonances with varying degrees of resonance splitting. For resonance R2 without significant splitting, as shown in Fig.~\ref{fig:2}(c), a "virtual" backward propagating mode can still be obtained through fitting. Consequently, our simulation is unaffected by varying degrees of resonance splitting and remains consistent across various scenarios, including the generation of separable states and entangled states.

\setcounter{equation}{0}
\renewcommand{\theequation}{C\arabic{equation}}
\section*{APPENDIX C: METHODS}
\subsection*{Purity Measurement}
\begin{figure}
\includegraphics[width=1\linewidth]{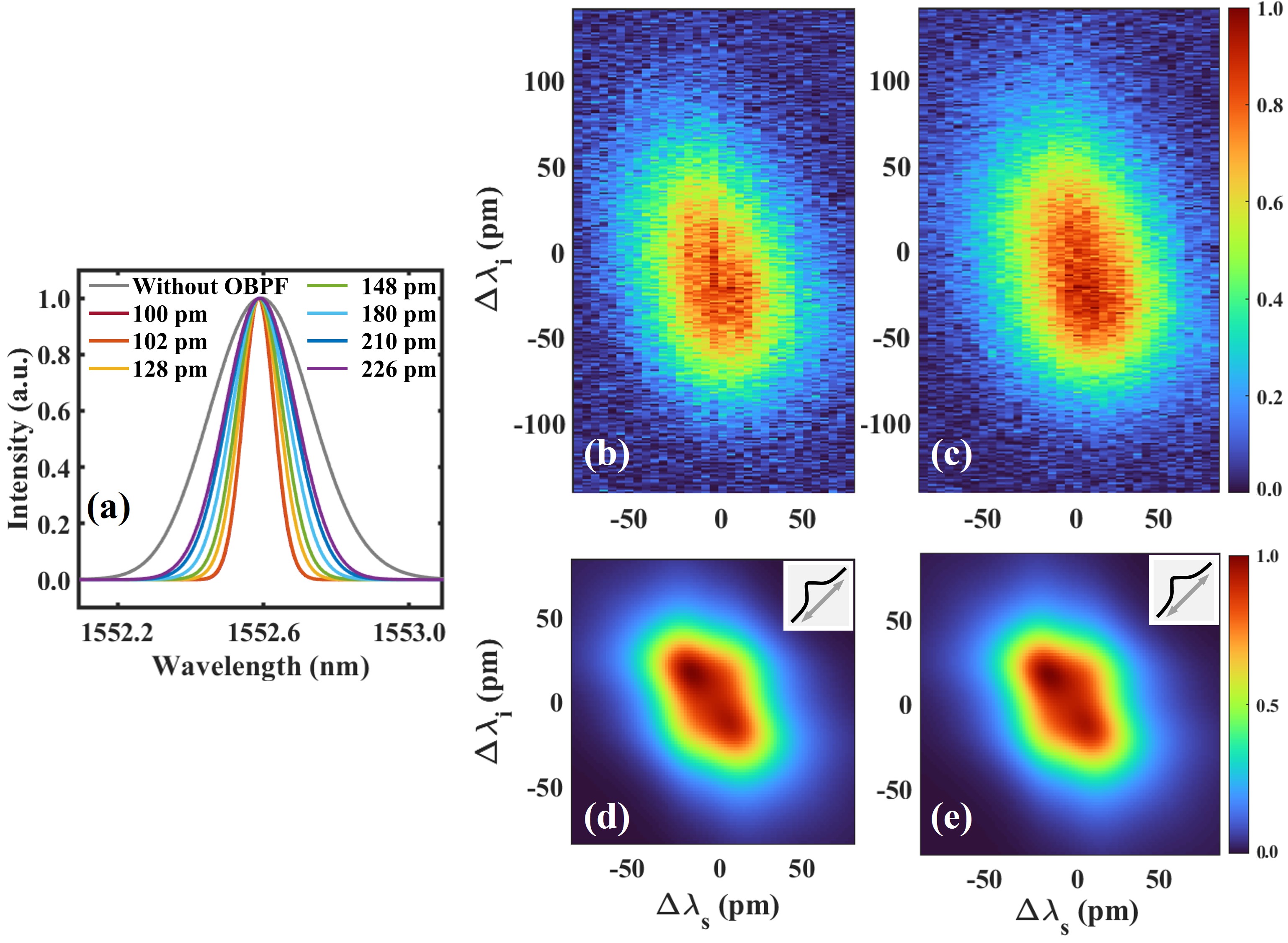}
\caption{\label{fig:8} (a) Measured spectra for pump pulses with various spectral FWHM and for the pulse without the OBPF filtering. (b)-(c) Measured JSIs with pump pulses of 210 pm (b) and 148 pm (c) FWHM, both consisting of data points in a 36 by 282 grid. (d) Modulus squared of the simulated JSA corresponding to (b). Inset: sketch of its ADP function. (e) Modulus squared of the simulated JSA corresponding to (c). Inset: sketch of its ADP function.}
\end{figure}

Figure~\ref{fig:8} shows the measured spectra of pulses generated by the pulsed laser and filtered by the OBPF with various bandwidths. For the $g_{u}^{\left( 2 \right)}\left( \Delta t \right)$ experiment in the separable state generation regime, the FWHM of the input pump pulses is set to 102, 128, 148, 180, 210, and 226 pm, respectively. The pulse with a 100 pm FWHM is set for the entangled state generation with the central wavelength shifted to 1552.61 nm. Spectral shapes are set to Gaussian-like for all experiments. The measured JSI with pump FWHM of 148 pm and 210 pm are shown in Figs.~\ref{fig:8}(b) and~\ref{fig:8}(c), respectively, with purity of $91.5\%$ and $93.9\%$ under the flat-phase assumption. The associated simulations yield purity values of $93.6\%$ and $94.6\%$, obtained from the simulated results shown in Figs.~\ref{fig:8}(d) and~\ref{fig:8}(f) under the flat-phase assumption. Because of pump resonance splitting, the bandwidth of the ADP functions (shown in the insets) does not change significantly as the pump's FWHM decreases from 226 pm to 148 pm. As a result, the purity of the generated single photons experiences a minor decrease.

\subsection*{Brightness Measurement}
The coupling loss for grating couplers (${\alpha}_{gc}$) is $7.0\pm 0.2$ dB per facet, while the insertion loss for DEMUXs (${\alpha}_{de}$) is $1.9\pm 0.1$ dB per stage. For signal/idler photons, the total single-channel loss (${\alpha}_{tot}$) from the bus waveguide output to the input of the SNSPD can be expressed as ${{\alpha }_{tot}}=1-{{10}^{({-{\alpha }_{gc}}-2{{\alpha }_{de}})/10}}$ (in percentage), neglecting losses from all fiber connections. The average detection efficiency ($\eta_d$) of the SNSPD (Photec) is $90\%$. The on-chip signal/idler counts ($C_X$, where $X=S,I$) and the on-chip coincidence counts (${C}_{CC}$) can be expressed using the formulas ${{C}_{X}}={{C}_{X-raw}}/{{\eta }_{d}}/\left( 1-{{\alpha }_{tot}} \right)$ and ${{C}_{CC}}={{C}_{CC-raw}}/\eta _{d}^{2}/{{\left( 1-{{\alpha }_{tot}} \right)}^{2}}$. Here, ${C}_{X-raw}$ and ${C}_{CC-raw}$ represent the raw single-channel and coincidence counts recorded by TDC, respectively. Besides, the DC and the ACC are estimated using the same method as the $C_X$ and ${C}_{CC}$, respectively.

\subsection*{MRR-based Photonic Differentiator}
The transfer function of an all-pass MRR is expressed as
\begin{equation}
\label{Eq:C1}
H\left( \omega  \right)=\frac{{{\tau }_{c}}-{{\alpha }_{rt}}\exp \left[ -j\omega {{T}_{s}} \right]}{1-{{\tau }_{c}}{{\alpha }_{rt}}\exp \left[ -j\omega {{T}_{s}} \right]},
\end{equation}
where ${\tau}_{c}$ is the self-coupling coefficient of the coupling region, ${\alpha}_{rt}$ is the round-trip loss coefficient, and ${{T}_{s}}$ is the round-trip time delay. A Nth-order optical temporal DIFF with the transfer function $H\left( \omega  \right)={{\left[ i\left( \omega -{{\omega }_{0}} \right) \right]}^{N}}$ approximates the MRR transfer function within a limited frequency range centered on the resonant frequency\cite{zheng2014fractional}. The order ${N}$ is determined by the MRR coupling condition.

	\bibliographystyle{unsrt}
 	\bibliography{document.bib}
 	
	\end{sloppypar}
\end{document}